\documentstyle[prl,aps,floats,epsf]{revtex}

\begin{document}
\draft

\twocolumn[\hsize\textwidth\columnwidth\hsize\csname
@twocolumnfalse\endcsname

\title{
Easy collective polarization switching in ferroelectrics
}

\author{A.M. Bratkovsky$^{1}$ and A.P. Levanyuk$^{1,2}$}
\address{$^{1}$Hewlett-Packard Laboratories, 1501 Page Mill Road, Palo Alto,
California 94304\\
$^{2}$Departamento de F\'{i}sica de la Materia Condensada, C-III,
Universidad Aut\'{o}noma de Madrid, 28049 Madrid, Spain
}
\date{May 18, 2000}
\maketitle

\begin{abstract}

The  actual mechanism of polarization switching in ferroelectrics remains 
a puzzle for many decades, since
the usually estimated barrier for nucleation and growth is insurmountable (``paradox
of the coercive field"). 
To analyze the mechanisms of the nucleation we consider the exactly solvable case
of a ferroelectric film with a ``dead" layer at the interface with electrodes.
The classical nucleation is easier in this case but still impossible, since the calculated
barrier is huge. We have found that the {\em interaction} between the nuclei is, 
however, long range,  hence one has to study an  {\em ensemble} of the nuclei. 
We show that there are the ensembles of small (embryonic) nuclei that grow 
{\em without the barrier}. We submit that the interaction between nuclei 
is the key point for solving the paradox.


\end{abstract}
\vskip 2pc ] 

The polarization switching in ferroelectrics (FEs) is most commonly used in
applications (capacitors, memory elements), yet this process remains the
least understood in spite of numerous experimental studies. As a rule, the
ferroelectrics are switching in the field $E_{c}$ which is some order of
magnitude lower than the so called ``thermodynamic coercive field'' $E_{c0}$%
\cite{landauer,janovec,kay,lem96,scott96}. The latter is the field at which
the homogeneously polarized ferroelectric loses its stability in external
field applied in opposite direction to the polarization. The fact that the
switching takes place well before the point of instability is reached means
that it proceeds by {\em inhomogeneous} nucleation and growth of {\em domains%
} of a new phase. But how the domains nucleate? The difficulty in answering
this question has been emphasized by Landauer in late 1950s \cite{landauer}.
His and later estimates \cite{janovec,kay} showed that the energy barrier
for creating a nucleus with reversed polarization is practically
insurmountable, $U^{\ast }\gtrsim 10^{3}k_{B}T$ in the field of about $100$%
kV/cm. This problem, or a ``paradox of a coercive field'', has in fact been
realized for domain nucleation in ferromagnets in 1938\cite{doring38}. The
experimental coercive field for the bulk magnetic materials was known to be
many times smaller than that suggested by the micromagnetics theory
(``Brown's paradox'') \cite{brown65,aharoni98}. The disagreement is
especially striking for hard magnetic materials, where the situation is
closer to the ferroelectrics considered in the present paper. There were
numerous suggestions over the years that some defects can assist the
nucleation and reduce the coercive field down to experimentally observed
values. The situation in magnetics was summarized by Brown in 1965 who noted
that the idea is plausible but ``there has been no strikingly successful
calculation based on a completely realistic model''\cite{brown65}. Since
then the situation has apparently remained the same\cite{aharoni98}, whereas
a switching in fine magnetic particles seems to be fairly described by
micromagnetics theory.

The goal of the present paper is to suggest a possible new mechanism for the
polarization switching in ferroelectric materials. To this end we consider
an exactly solvable model of a ferroelectric material with an extended
inhomogeneity of its dielectric response at the ferroelectric-electrode
interface, which is called a ``passive'' or ``dead'' layer (for references
see e.g.\cite{prl,lem96,royt99}).

The main feature of the ferroelectric film with the dead layer is that the
film exists in a polydomain state at any thickness of the dead layer \cite
{prl}. The monodomain state in this system, which can be produced by cooling
in the field, would tend to transform into the equilibrium polydomain state.
The transformation is {\em favored} by the fact that the polydomain state
reduces the energy of the depolarizing field in spite of increased surface
energy of the domain walls. This is contrary to the usual notion that the
depolarizing field hinders the nucleation\cite{landauer}, yet the classical
nucleation remains impossible.

First, we shall study the classical nucleation of the individual domains of
the new phase. The nuclei will be assumed below to have a form of stripes or
cylinders with the domain walls perpendicular to the plane of the film ($c$%
-domains), which is a reasonable approximation. We shall evaluate the
barrier for their nucleation and show that for the individual nucleus it is
practically insurmountable, although the dead layer helps to reduce it. We
then abandon the classical approach and study the {\em interaction} between
nuclei and find that it is long range.

It becomes clear that when an individual nucleus cannot grow the {\em %
ensemble of nuclei} may be able to. As an example of such an ensemble we
consider a periodic array of nuclei and show that it indeed provides a path
to the equilibrium state. We shall show that there is {\em no energy barrier 
}for its growth already when the nuclei become larger than the thickness of
the domain wall $W$. After this `embryonic' state has been passed the free
energy of the system decreases monotonously as the nuclei grow. Obviously,
the energy of embryonic nuclei is much smaller than the critical energy of
the Landauer's nucleus\cite{landauer}. This is simply related to the fact
that the size of the critical Landauer needle-like nucleus is large, the
radius of its base is $r_{c}^{\text{L}}=WE_{c0}/E_{\text{ext}}\gg W,$ where $%
E_{\text{ext}}$ $\left( \ll E_{c0}\right) $ is the external field$,$
therefore its energy $U_{\text{L}}^{\ast }$ is huge. In the present case of
the dead layer $E_{\text{ext}}$ should be replaced by the (small)\
depolarizing field in the ferroelectric and the critical radius of the
Landauer's nucleus remains $\gg W.$ Since for the collective nucleation $%
r_{c}^{\text{coll}}\sim W\ll r_{c}^{\text{L}},$ the incurred energy barrier
for a nuclei to grow beyond the embryonic state $U_{\text{coll}}^{\ast },$
if any, should be much smaller than the standard barrier $U_{\text{L}}^{\ast
}$ for individual nuclei. The equilibrium density of the embryos is large
(see below) and the nucleation of the macroscopic domain (and the coercive
field)\ would be determined by the waiting time for optimal fluctuation and
its dependence on the electric field for an ensemble which can grow without,
or almost without, the barrier, 

We shall mainly discuss the energy aspects of the nucleation. What the
present analysis demonstrates is that the nucleation of ferroelectric
domains is facilitated by an interaction between the nuclei. Note that no
such interaction is taken into account within the Kolmogorov-Avrami\ theory 
\cite{kolmogor} which is widely used to treat switching in ferroelectrics.

We shall consider first the problem of the barrier for single stripe and
then cylindrical domains, which appear to be insurmountable. We establish,
however, that the interaction between the nuclei is long range and may give
a clue to actual nucleation process. We then turn to exactly solvable case
of an {\em ensemble} of stripe domains and show that the barrier for
nucleation is actually {\em zero} already when the nuclei are at the
embryonic stage.

The geometry of the present problem for ferroelectric with dead layer is
illustrated in Fig.~1. For a short-circuited electrodes (zero bias voltage)
the free energy of the system is $\widetilde{F}=F_{0}+U_{es},$ where the
electrostatic energy $U_{es\text{ }}$is \cite{prl}:

\begin{equation}
U_{es}=\frac{1}{2}\int_{{\rm FE}}\text{d}{\cal A}\sigma \varphi 
\label{Ues0}
\end{equation}
where $\sigma $ is the bound charge due to the spontaneous polarization, $%
\varphi $ is the electrostatic potential, while $F_{0}$ includes the surface
energy of the domain walls, and the integration goes over the FE\ surface.
The electrostastic potential $\varphi $ is found from solving the Poisson
equation for assumed domain structure \cite{prl}. We obtain with the use of
the Fourier transformation the total electrostatic energy for arbitrary
one-dimensional domain structure as 
\begin{equation}
U_{es}=2\int_{-\infty }^{\infty }\frac{\text{d}k}{k}\frac{\mid \sigma
_{k}\mid ^{2}}{\sqrt{\varepsilon _{a}\varepsilon _{c}}\coth \left( \sqrt{%
\frac{\varepsilon _{a}}{\varepsilon _{c}}}\frac{kl}{2}\right) +\varepsilon
_{g}\coth \frac{kd}{2}},  \label{Ues}
\end{equation}
where $\sigma _{k}\equiv \int_{-\infty }^{\infty }dx\exp (-ikx)\sigma
(x,z=l/2)$ is the Fourier component of the surface bound charge $\sigma
(x,z=l/2),$ $l$ the thickness of the FE film with $\varepsilon _{c(a)}$ the
dielectric constants in $c$ $(a)$ direction, $d$ the thickness of the dead
layer (Fig. 1).

We begin with the case of a nucleus in the center of the plate, i.e. with
the following distribution of the bound charge 
\begin{eqnarray}
\sigma (x,z &=&+(-)l/2)=-(+)P_{s},\qquad \left| x\right| <a/2;  \nonumber \\
\sigma (x,z &=&+(-)l/2)=+(-)P_{s},\qquad a/2<\left| x\right| <R,
\label{sigmaofx}
\end{eqnarray}
where $2R$ is the width of the plate and $a$ is the width of the nucleus,
and $P_{s}$ the spontaneous polarization. In this case $\sigma _{k}=\frac{%
2P_{s}}{k}\left( -2\sin \frac{ka}{2}+\sin kR\right) .$ The dependence of all
physical quantities on $R$ disappears in a limit $R\rightarrow \infty $, as
it should. It is handy to always subtract the (constant)\ electrostatic
energy of the uniformly polarized sample, which is characterized by the
Fourier transform of the bound charge $\bar{\sigma}_{k}=(2P_{s}/k)\sin kR.$
By using Eq.(\ref{sigmaofx}) one then finds the change of the electrostatic
energy due to creation of a {\em stripe} nucleus 
\begin{eqnarray}
U^{\rm stripe}_{es} &=&-8\pi \varepsilon _{g}^{-1}P_{s}^{2}da\left[ 1-\frac{a}{\pi d}\ln
\left( \frac{e^{3/2}d}{a}\right) \right] ,\qquad a\lesssim d;
\label{centra>d} \\
&=&-8\varepsilon _{g}^{-1}P_{s}^{2}d^{2}\ln \left( \frac{e^{3/2}a}{d}\right)
,\qquad a\gg d;
\end{eqnarray}
for $d<l\sqrt{\varepsilon _{c}/\varepsilon _{a}},$ $\varepsilon _{g}=\sqrt{%
\varepsilon _{a}\varepsilon _{c}}$. Note that this is the {\em change} in
electrostatic energy with respect to uniformly polarized sample when the
nucleus is present, so it does not apply to completely reversed sample. The
electrostatic energy {\em favors} the nucleation, since the domains reduce
the energy of stray field . The total energy of the stripe nucleus per unit
length is 
\begin{equation}
\tilde{F}_{\text{stripe}}(a)=2l\gamma +U_{es}-2P_{s}E_{\text{ext}}al,
\label{Ftot}
\end{equation}
where $\gamma =P_{s}^{2}\Delta $ is the surface energy of the domain wall
with $\Delta $ the temperature dependent characteristic length\cite{prl}. We
see that the gain in electrostatic energy eventually overwhelms the surface
energy, and there appears an exponentially wide barrier for the nucleus in $%
\tilde{F}(a)$ when $E_{{\rm ext}}=0.$

We shall now consider the case of {\em two} nuclei of a new phase in a form
of stripe domains. With the help of Eq.$~\left( \ref{Ues}\right) $ one can
easily calculate the change of the electrostatic energy due to formation of
two nuclei having the same width $a$, with the separation $r$\ between their
centers. We find for the energy of the interaction of two stripe domains per
unit length for $r\gg a$
\begin{equation}
U^{\rm stripe}_{int}\left( r\right) =2\varepsilon
_{g}^{-1}P_{s}^{2}a^{2}d^{2}\frac{1}{%
r^{2}},  \label{eq:UintStr}
\end{equation}
which corresponds to long range dipole-dipole interaction between two
stripes \cite{dipole}. This observation indicates that the interactions in a
system with an ensemble of nuclei would be very important. Note that the
interaction$\;$of the stripe nucleus with the edge of the sample $U_{\text{%
{\rm edge}}}$ is also long range, $U_{\text{{\rm edge}}}(x_{0})=2\varepsilon
_{g}^{-1}P_{s}^{2}ad^{2}/x_{0},$ and they are repelled from the edge.

Similar treatment can be repeated for a single and a pair of cylindrical
nuclei. The expression for the electrostatic energy for cylindrical nucleus
is similar to that for the stripe case (\ref{Ues}) 
\begin{equation}
U_{es}=2\int_{0}^{\infty }\text{d}k\frac{\mid \sigma _{k}\mid ^{2}}{\sqrt{%
\varepsilon _{a}\varepsilon _{c}}\coth \left( \sqrt{\frac{\varepsilon _{a}}{%
\varepsilon _{c}}}\frac{kl}{2}\right) +\varepsilon _{g}\coth \frac{kd}{2}},
\label{Uesc}
\end{equation}
where for the nucleus with the radius $a$ in the center of the slab with the
radius $R$ we obtain $\sigma _{k}\equiv 2\pi \int_{0}^{R}$d$%
rrJ_{0}(kr)\sigma (r,z=l/2)=(2\pi P_{s}/k)\left[ RJ_{1}(kR)-2aJ_{1}(ka)%
\right] ,$ while for the uniformly polarized sampe $\bar{\sigma}_{k}=(2\pi
P_{s}/k)RJ_{1}(kR),$ with $J_{n}(z)$ the Bessel function. The integral in
the expression (\ref{Uesc}) for the electrostatic energy of the {\em %
cylindrical} nucleus can be evaluated with the result 
\begin{eqnarray}
U^{\rm cyl}_{es} &=&-8\pi ^{2}\varepsilon _{g}^{-1}P_{s}^{2}da^{2},\qquad a\lesssim d;
\nonumber \\
&=&-8\pi \varepsilon _{g}^{-1}P_{s}^{2}d^{2}a\ln \frac{8a}{e^{1/2}d},\qquad
a\gg d;  \label{Uescyl}
\end{eqnarray}
The free energy of one cylindrical nucleus is 
\begin{equation}
\tilde{F}_{\text{cyl}}(a)=2\pi al\gamma +U_{es}-2\pi P_{s}E_{\text{ext}%
}la^{2}.  \label{Fcyl}
\end{equation}
It is obvious from Eqs.~(\ref{Uescyl}),(\ref{Fcyl}) that the gain in the
electrostatic energy eventually overwhelms the growth of the the surface
energy of the domain wall with increase of \ the radius $a$ of the nucleus.
The critical radius is exponentially large when $E_{{\rm ext}}=0$, it
behaves roughly as $a_{c}\sim d\exp \left( 0.3a_{K}^{2}/d^{2}\right) ,$
where $a_{K}=\left( 0.3\tilde{\varepsilon}\Delta l\right) ^{1/2}$ is the
Kittel period, $\tilde{\varepsilon}=\varepsilon _{g}+\sqrt{\varepsilon
_{a}\varepsilon _{c}}$\cite{prl}. The corresponding barrier height $\tilde{F}%
_{c}$ is very large, $\tilde{F}_{c}\gg P_{s}^{2}d_{at}^{3}\sim E_{at},$
where $d_{at}$ is the characteristic ``atomic'' length (of the order of the
lattice parameter) and the ``atomic'' energy $E_{at}$ amounts to a few eV.
Thus, in this case the barrier is also huge, comparable to the barrier for
the Landauer's nucleus, and its growth is prohibitively expensive.

The interaction between two {\em cylindrical} nuclei can be estimated for $%
r\gg a$ as 
\begin{equation}
U^{\rm cyl}_{int}\left( r\right) =16\pi ^{2}\varepsilon
_{g}^{-1}P_{s}^{2}a^{4}d^{2}%
\frac{1}{r^{3}}  \label{eq:UintCyl}
\end{equation}
In this case it appears to also be long range of a dipole-dipole type \cite
{dipole}.

The importance of these interactions lies in their long range rather than
sign. Therefore, one has to accurately evaluate the total energy for a
system of nuclei, since it does not reduce to a sum of asymptotic
interactions (\ref{eq:UintStr}) or (\ref{eq:UintCyl}). The long range
interactions give us a clue to the mechanism of nucleation and growth of a
new phase. To illustrate that it indeed may solve the problem, we shall
consider a system of stripe-like domains. The results below will demonstrate
that the electrostatic energy not only favors nucleation but {\em eliminates}
an energy barrier for the growth of nuclei in an {\em ensemble} when their
size is larger than the domain wall width, i.e. the switching proceeds
collectively.
\begin{figure}[th]
\epsfxsize=3.4in
\epsffile{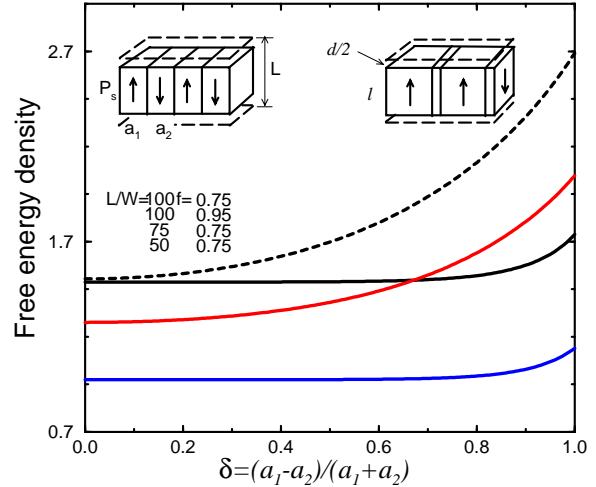}
\vspace{.02in}
\caption{ Free energy density $\tilde{F}/{\cal A}P_{s}^{2}$ for the growth
of ensemble of nuclei for different values $L/W$ and the fraction of volume
occupied by the ferroelectric $f=l/L$ ($d=L-l$ is the thickness of the dead
layer). Note that there is {\em no barrier} for growth of very small nuclei
(from $\protect\delta \approx 1$) towards the domain pattern with equal
width $\protect\delta =0$ of the domains of both phases $P_{s}$ and $-P_{s}$
(Kittel state). The flatness of the free energy $\tilde{F}$ in the vicinity
of the Kittel state demonstrates remarkable softness of the ferroelectric
with the dead layer. Inset shows schematics of the growth of the ensemble of
nuclei (narrow domains of the new phase). }
\label{fig:1}
\end{figure}

To study the {\em collective} nucleation we shall analyze a domain structure
with stripe domains of opposite polarization of widths $a_{1}$ and $a_{2}$,
the period $T=a_{1}+a_{2},$ and the asymmetry parameter $\delta =\frac{%
a_{1}-a_{2}}{a_{1}+a_{2}},$ which measures a net polarization of the film 
\cite{prl}. For zero external bias voltage assumed throughout the present
paper $\delta =0$ in \ the equilibrium (polydomain) state, whereas in the
monodomain state $\delta =1.$ At $\delta \rightarrow 1$ the system consists
of very narrow domains with the polarization opposite to the net
polarization (Fig. 1, inset), i.e. it is a periodic ensemble of nuclei. The
free energy $\tilde{F}$ of this system at {\em zero} external bias is given
by

\begin{eqnarray}
\frac{\tilde{F}}{{\cal A}P_{s}^{2}} &=&\frac{2\pi d\delta ^{2}}{\varepsilon
_{c}\left( d/l\right) +\varepsilon _{g}}+\frac{2\Delta l}{T}+\frac{16T}{\pi
^{2}}\sum_{j=0}^{\infty }\frac{1}{\left( 2j+1\right) ^{3}}\frac{1}{D_{2j+1}}
\nonumber \\
&&+\frac{8T}{\pi ^{2}}\sum_{n=1}^{\infty }\left( -1\right) ^{n}\frac{1-\cos
\pi n\delta }{n^{3}}\frac{1}{D_{n}},  \label{eq:Fdel}
\end{eqnarray}
where $D_{n}=\sqrt{\varepsilon _{a}\varepsilon _{c}}\coth \sqrt{\frac{%
\varepsilon _{a}}{\varepsilon _{c}}}\frac{\lambda _{n}l}{2}+\varepsilon
_{g}\coth \frac{\lambda _{n}d}{2}$\ and $\lambda _{n}=2\pi n/T$ \cite{prl}.
Fortunately, this free energy can be found analytically in the asymptotic
case of narrow dead layer $(d\ll l)$ and correspondingly wide period of the
domain structure, $T\gg d,$ when $\varepsilon _{g}=\sqrt{\varepsilon
_{a}\varepsilon _{c}}.$ We have found earlier that the period of the domain
structure in this case is exponentially large, $T=0.95d\exp
(0.4a_{K}^{2}/d^{2})$. We note that in this case $D_{n}=\varepsilon
_{g}\left( 1+\coth \frac{\lambda _{n}d}{2}\right) $ and summation in $\left( 
\text{\ref{eq:Fdel}}\right) $ can be performed to yield 
\begin{eqnarray}
\frac{\tilde{F}}{{\cal A}P_{s}^{2}} &=&\frac{2\pi d\delta ^{2}}{\varepsilon
_{c}\left( d/l\right) +\varepsilon _{g}}+\frac{2\Delta l}{T}+\frac{4T}{\pi
^{2}\varepsilon _{g}}[\zeta (3)-2Li_{3}\left( e^{-b}\right)   \nonumber \\
&&+\frac{1}{4}Li_{3}\left( e^{-2b}\right) -Li_{3}\left( -e^{-b}\right)  
\nonumber \\
&&-%
\mathop{\rm Re}%
Li_{3}\left( -e^{i\pi \delta }\right) +%
\mathop{\rm Re}%
Li_{3}\left( -e^{i\pi \delta -b}\right) ],  \label{Faexact}
\end{eqnarray}
where $b=4\pi d/T\ll 1,$ and $Li_{n}(z)\equiv \sum_{k=1}^{\infty }z^{k}/k^{n}
$. For the case of an ensemble of narrow domains, $\delta \simeq 1$ the free
energy can be found from the known asymptotic behavior of the $Li_{n}(z)$
function \cite{Ryzhik}, yielding an approximate expression 
\begin{equation}
\frac{\tilde{F}}{{\cal A}P_{s}^{2}}=\frac{2\pi d\delta ^{2}}{\varepsilon
_{c}\left( d/l\right) +\varepsilon _{g}}+\frac{2\Delta l}{T}+\frac{T}{%
\varepsilon _{g}}\left( 1-\delta \right) ^{2}\ln \frac{e^{3}b^{2}}{\pi
^{2}(1-\delta )^{2}}.  \label{Fas}
\end{equation}

Now everything depends on how the free energy $\tilde{F}$ behaves as a
function of $\delta $ when the nuclei grow, i.e. when $\delta $ reduces from
unity towards zero. One can easily see that at $x\equiv 1-\delta \ll 1$ the
free energy $\tilde{F}/{\cal A}P_{s}^{2}$ is given, with respect to a
constant, by the function 
\begin{equation}
f(x)\equiv -\frac{4\pi dx}{\varepsilon _{c}\left( d/l\right) +\varepsilon
_{g}}+\frac{T}{\varepsilon _{g}}x^{2}\ln \frac{e^{3}b^{2}}{\pi ^{2}x^{2}},
\label{fx}
\end{equation}
where $f(0)=0.$ This function does {\em not} have a barrier as a function of 
$x$ when $T$ is kept constant. Indeed, the second term in (\ref{fx}) has an
exponentially small maximum ($=T\varepsilon _{g}^{-1}x_{0}^{2}$) at $%
x_{0}=4ed/T\ll 1.$ It is, however, suppressed by the first term, which
corresponds to the energy of homogeneous field created by the net
polarization and is linear in $x.$ As a result, there appears to be {\em no
barrier} for the growth of nuclei. Note that we have actually restricted the
system's path for nucleation by constraining the domain pattern to a fixed
period. Even under this constraint the growth proceeds without the energy
barrier, and this would be even more so if we were to lift the constraint
and allow the system to follow an optimal path to equilibrium. This behavior
does not depend on the approximation we have made for evaluating the free
energy, the exact calculation of the free energy (\ref{eq:Fdel})\ for all $%
\delta $ shows that the {\em collective} growth of nuclei we just described
proceeds {\em without the} {\em barrier} (Fig. 1). As mentioned above, the
smallest size of nuclei where the present analysis applies is of the order
of the domain wall width $W$ and the barrier for the nucleation of such
small embryonic nuclei is expected to be zero or much smaller than the usual
estimates\cite{landauer,janovec,kay} for individual nucleation.

For nucleation to proceed by the present mechanism the only condition is the
presence of extended dielectric inhomogeneity in a sample. The external
field does promote growth of the nuclei, but the nucleation in the present
system occurs even without it. The likely requirement is that the lateral
extent of this inhomogeneity should be much larger than the period of the
equilibrium domain (Kittel)\ structure $a_{K}.$ One can estimate the rate of
embryo nucleation in $1{\rm cm}^{3}$ as $\sim (1/W^{3}\tau _{ph})\exp
(-U_{em}/kT),$ where $\tau _{ph}$ is the characteristic (optical phonon)
time, and the ``atomic'' estimate of the energy of the embryo is $U_{em}\sim
\gamma W^{2}\sim E_{at}\sqrt{T_{c}/T_{at}}\sim 2-3\cdot 10^{3}{\rm K}\cite
{embryo},$ with the characteristic ``atomic'' temperature $T_{at}\sim 10^{4}%
{\rm K}$. Taking a conservative estimate of the embryo lifetime as $\sim
10\tau _{ph},$ one finds the equilibrium density of the embryos $\sim 10^{17}
$ cm$^{-3}.$ At such high densities the embryos should `feel' the field of
each other, and favorable ensembles should appear within a reasonable time,
unlike in the case of the Landauer's nucleus where the expectation time much
exceeds the lifetime of the universe.

In conclusion, we have suggested a possible way to solve the ``{\it paradox
of the coercive field''} by demonstrating a collective mode of domain growth
past the embryonic stage with sizes about the domain wall thickness $W$,
which proceeds without energy barrier. The origin of this cooperative
phenomenon is a long-range interaction of electrostatic origin between the
nuclei. The possible screening by free charges in ferroelectric does not
seem to be important, since the conductivity of ferroelectrics is usually
too low to have any effect \cite{landauer}. As a corollary, we note that the
Kolmogorov-Avrami model (KA)\ \cite{kolmogor} is inapplicable to growth of
domains in ferroelectrics, since the essential long-range interaction
between nuclei is completely neglected in this approach (note that KA fully
bypasses the question of how the domains were nucleated in the first place).
The present results are general, and have the implication that in
ferroelastic materials and possibly also magnetic materials the nucleation
would be facilitated by the long-range interaction between nuclei of a new
phase. This could alleviate the Brown's paradox of the coercive field in
relation to switching in bulk ferromagnets.

We acknowledge helpful discussions with A. Aharoni, V.V. Osipov, A.L.
Roytburd, and A.K. Tagantsev.


\begin{references}
\bibitem{dipole}  Note that the stripe domain and its mirror image form the
dipole moment $2P_{s}ad$ (per unit length). For cylindrical nucleus it
equals $2P_{s}\pi a^{2}d.$ The coefficients in Eqs.(\ref{eq:UintStr}),(\ref
{eq:UintCyl}) reflect the interaction between infinite stacks of the mirror
images.
\bibitem{embryo}  At these small sizes it does not make sense to consider a
particular geometry of embryos, so we make the order of magnitude estimates.
\end{references}
\end{document}